\documentstyle[amsfonts,12pt]{article}
\newtheorem{th}{Theorem}[section]
\newtheorem{prop}[th]{Proposition}

\newtheorem{lem}[th]{Lemma}

\begin{document}
\title{A basic inequality for submanifolds in a cosymplectic space form}
\author{Jeong-Sik Kim and Jaedong Choi\thanks{
The authors were partially supported by Korea Science and Engineering Foundation Grant
(R01-2001-00003).}}
\maketitle
\begin{abstract}
For submanifolds tangent to the structure vector field in cosymplectic space
forms, we establish a basic inequality between the main intrinsic invariants
of the submanifold, namely its sectional curvature and scalar curvature on
one side; and its main extrinsic invariant, namely squared mean curvature on
the other side. Some applications including inequalities between the
intrinsic invariant $\delta _{M}$ and the squared mean curvature are given.
The equality cases are also discussed. \bigskip
\end{abstract}

\section{Introduction}

To find simple relationships between the main extrinsic invariants and the
main intrinsic invariants of a submanifold is one of the natural interests
in the submanifold theory. Let $M$ be an $n$-dimensional Riemannian
manifold. For each point $p\in M$, let $\left( \inf K\right) \left( p\right)
$ $=$ $\inf \left\{ K\left( \pi \right) :\mbox{plane sections\ }\pi \subset
T_{p}M\right\} $. Then, the well defined intrinsic invariant $\delta _{M}$
for a $M$ introduced by B.-Y. Chen([4]) is
\begin{equation}
\delta _{M}\left( p\right) =\tau \left( p\right) -\left( \inf K\right)
\left( p\right) ,  \label{sigma-inv}
\end{equation}
where $\tau $ is the scalar curvature of $M$ (see also [6]).
\medskip

In [3], Chen established the following basic inequality involving
the intrinsic invariant $\delta _{M}$ and the squared mean curvature for $n$%
-dimensional submanifolds $M$ in a real space form $R\left( c\right) $ of
constant sectional curvature $c$:
\begin{equation}
\delta _{M}\leq \frac{n^{2}\left( n-2\right) }{2\left( n-1\right) }\left\|
H\right\| ^{2}+\frac{1}{2}\left( n+1\right) \left( n-2\right) c.
\label{chen-in-1}
\end{equation}
The above inequality is also true for anti-invariant submanifolds in complex
space forms $\widetilde{M}\left( 4c\right) $ as remarked in [7]. In
[5], he proved a general inequality for an arbitrary submanifold
of dimension greater than two in a complex space form. Applying this
inequality, he showed that (\ref{chen-in-1}) is also valid for arbitrary
submanifolds in complex hyperbolic space $CH^{m}\left( 4c\right) $. He also
established the basic inequality for a submanifold in a complex projective
space $CP^{m}$. \medskip

A submanifold normal to the structure vector field $\xi $ of a contact
manifold is anti-invariant. Thus $C$-totally real submanifolds in a Sasakian
manifold are anti-invariant, as they are normal to $\xi $. An inequality
similar to (\ref{chen-in-1}) for $C$-totally real submanifolds in a Sasakian
space form $\tilde{M}\left( c\right) $ of constant $\varphi $-sectional
curvature $c$ is given in [8]. In [9], for submanifolds in
a Sasakian space form $\tilde{M}(c)$ tangential to the structure vector
field $\xi $, a basic inequality along with some applications are presented.
\medskip

There is another interesting class of almost contact metric manifolds,
namely cosymplectic manifolds([10]). In this paper, submanifolds
tangent to the structure vector field $\xi $ in cosymplectic space forms are
studied. Section~\ref{prel-lc-chen1} contains necessary details about
submanifolds and cosymplectic space forms are given for further use. In
section~\ref{basic-lc-chen1}, for submanifolds tangent to the structure
vector field $\xi $ in cosymplectic space forms, we establish a basic
inequality between the main intrinsic invariants, namely its sectional
curvature function $K$ and its scalar curvature function $\tau $ of the
submanifold on one side, and its main extrinsic invariant, namely its mean
curvature function $\left\| H\right\| $ on the other side. In the last
section, we give some applications including inequalities between the
intrinsic invariant $\delta _{M}$ and the extrinsic invariant $\left\|
H\right\| $. We also discuss the equality cases.

\section{Preliminaries\label{prel-lc-chen1}}

Let $\tilde{M}$ be a $\left( 2m+1\right) $-dimensional almost contact
manifold([2]) endowed with an almost contact structure $\left(
\varphi ,\xi ,\eta \right) $, that is, $\varphi $ is a $\left( 1,1\right) $
tensor field, $\xi $ is a vector field and $\eta $ is $1$-form such that $%
\varphi ^{2}=-I+\eta \otimes \xi $ and $\eta \left( \xi \right) =1$. Then, $%
\varphi \left( \xi \right) =0$ and $\eta \circ \varphi =0$. \medskip

Let $g$ be a compatible Riemannian metric with $\left( \varphi ,\xi ,\eta
\right) $, that is, $g\left( \varphi X,\varphi Y\right) $ $=$ $g\left(
X,Y\right) $ $-$ $\eta \left( X\right) \eta \left( Y\right) $ or
equivalently, $g\left( X,\varphi Y\right) =-g\left( \varphi X,Y\right) $ and
$g\left( X,\xi \right) =\eta \left( X\right) $ for all $X,Y\in T\tilde{M}$.
Then, $\tilde{M}$ becomes an almost contact metric manifold equipped with an
almost contact metric structure $\left( \varphi ,\xi ,\eta ,g\right) $. An
almost contact metric manifold is {\em cosymplectic}([2]) if $%
\tilde{\nabla}_{X}\varphi =0$, where $\tilde{\nabla}$ is the Levi-Civita
connection of the Riemannian metric $g$. From the formula $\tilde{\nabla}%
_{X}\varphi =0$ it follows that $\tilde{\nabla}_{X}\xi =0$.

A plane section $\sigma $ in $T_{p}\tilde{M}$ of an almost contact metric
manifold $\tilde{M}$ is called a $\varphi $-{\em section} if $\sigma \perp
\xi $ and $\varphi \left( \sigma \right) =\sigma $. $\tilde{M}$ is of {\em %
constant} $\varphi $-{\em sectional curvature} if the sectional curvature $%
\tilde{K}(\sigma )$ does not depend on the choice of the $\varphi $-section $%
\sigma $ of $T_{p}\tilde{M}$ and the choice of a point $p\in $ $\tilde{M}$.
A cosymplectic manifold $\tilde{M}$ is of constant $\varphi $-sectional
curvature $c$ if and only if its curvature tensor $\tilde{R}$ is of the form([10])
\begin{eqnarray}
4\tilde{R}\left( X,Y,Z,W\right) &=&c\left\{ g\left( X,W\right) g\left(
Y,Z\right) -g\left( X,Z\right) g\left( Y,W\right) \right.  \nonumber \\
&&\left. +g\left( X,\varphi W\right) g\left( Y,\varphi Z\right) -g\left(
X,\varphi Z\right) g\left( Y,\varphi W\right) \right.  \nonumber \\
&&\left. -2g\left( X,\varphi Y\right) g\left( Z,\varphi W\right) \right.
\nonumber \\
&&\left. -g\left( X,W\right) \eta \left( Y\right) \eta \left( Z\right)
+g\left( X,Z\right) \eta \left( Y\right) \eta \left( W\right) \right.
\nonumber \\
&&\left. -g\left( Y,Z\right) \eta \left( X\right) \eta \left( W\right)
+g\left( Y,W\right) \eta \left( X\right) \eta \left( Z\right) \right\} .
\label{sect-curv}
\end{eqnarray}

Let $M$ be an $\left( n+1\right) $-dimensional submanifold of a manifold $%
\tilde{M}$ equipped with a Riemannian metric $g$. The Gauss and Weingarten
formulae are given respectively by $\tilde{\nabla}_{X}Y=\nabla _{X}Y+h\left(
X,Y\right) $ and $\tilde{\nabla}_{X}N=-A_{N}X+\nabla _{X}^{\perp }N$ for all
$X,Y\in TM$ and $N\in T^{\perp }M$, where $\tilde{\nabla}$, $\nabla $ and $%
\nabla ^{\perp }$ are respectively the Riemannian, induced Riemannian and
induced normal connections in $\tilde{M}$, $M$ and the normal bundle $%
T^{\perp }M$ of $M$ respectively, and $h$ is the second fundamental form
related to the shape operator $A$ by $g\left( h\left( X,Y\right) ,N\right)
=g\left( A_{N}X,Y\right) $. \medskip

Let $\left\{ e_{1},...,e_{n+1}\right\} $ be an orthonormal basis of the
tangent space $T_{p}M$. The mean curvature vector $H\left( p\right) $ at $%
p\in M$ is
\begin{equation}
H\left( p\right) \equiv \frac{1}{n+1}\sum_{i=1}^{n+1}h\left(
e_{i},e_{i}\right) .  \label{mean-curv}
\end{equation}
The submanifold $M$ is {\em totally geodesic} in $\tilde{M}$ if $h=0$, and
{\em minimal} if $H=0$. We put
\[
h_{ij}^{r}=g\left( h\left( e_{i},e_{j}\right) ,e_{r}\right) \quad \mbox{and}%
\quad \left\| h\right\| ^{2}=\sum_{i,j=1}^{n+1}g\left( h\left(
e_{i},e_{j}\right) ,h\left( e_{i},e_{j}\right) \right) .
\]

\section{A {\bf basic inequality\label{basic-lc-chen1} }}

Let $M$ be a submanifold of an almost contact metric manifold. For $X\in TM$%
, let
\[
\varphi X=PX+FX,\qquad PX\in TM,\ FX\in T^{\perp }M.
\]
Thus, $P$ is an endomorphism of the tangent bundle of $M$ and satisfies
\[
g\left( X,PY\right) =-g\left( PX,Y\right) ,\qquad X,Y\in TM.
\]
For a plane section $\pi \subset T_{p}M$ at a point $p\in M$,
\[
\alpha (\pi )=g\left( e_{1},Pe_{2}\right) ^{2}\qquad \mbox{and\qquad }\beta
(\pi )=(\eta (e_{1}))^{2}+(\eta (e_{2}))^{2}
\]
are real numbers in the closed unit interval $\left[ 0,1\right] $, which are
independent of the choice of the orthonormal basis $\left\{
e_{1},e_{2}\right\} $ of $\pi $. \medskip

We recall the following lemma from([3]).

\begin{lem}
\label{chen-lemma}If $a_{1},\ldots ,a_{n+1},a$ are $n+2$ $\left( n\geq
1\right) $ real numbers such that
\[
\left( \sum_{i=1}^{n+1}a_{i}\right) ^{2}=n\left(
\sum_{i=1}^{n+1}a_{i}^{2}+a\right) ,
\]
then $2a_{1}a_{2}\geq a$, with equality holding if and only if $%
a_{1}+a_{2}=a_{3}=\cdots =a_{n+1}$.
\end{lem}

Now, we prove the following

\begin{th}
\label{bi-th-1}Let $M$ be an $\left( n+1\right) $-dimensional $\left( n\geq
2\right) $ submanifold isometrically immersed in a $\left( 2m+1\right) $%
-dimensional cosymplectic space form $\tilde{M}\left( c\right) $ such that
the structure vector field $\xi $ is tangential to $M$. Then, for each point
$p\in M$ and each plane section $\pi \subset T_{p}M$, we have
\begin{equation}
\tau -K\left( \pi \right) \leq \frac{\left( n+1\right) ^{2}\left( n-1\right)
}{2n}\left\| H\right\| ^{2}+\frac{c}{8}\left( 3\left\| P\right\|
^{2}-6\alpha \left( \pi \right) +2\beta \left( \pi \right) +\left(
n+1\right) \left( n-2\right) \right) .  \label{bi}
\end{equation}
The equality in $\left( \ref{bi}\right) $ holds at $p\in M$ if and only if
there exists an orthonormal basis $\left\{ e_{1},\ldots ,e_{n+1}\right\} $
of $T_{p}M$ and an orthonormal basis $\left\{ e_{n+2},\ldots
,e_{2m+1}\right\} $ of $T_{p}^{\perp }M$ such that {\bf (a)} $\pi =
Span\left\{ e_{1},e_{2}\right\} $ and {\bf (b)} the forms of shape
operators $A_{r}\equiv A_{e_{r}},$ $r=n+2,\ldots ,2m+1$, become
\begin{equation}
A_{n+2}=\left(
\begin{array}{ccc}
\lambda & 0 & 0 \\
0 & \mu & 0 \\
0 & 0 & \left( \lambda +\mu \right) I_{n-1}
\end{array}
\right) ,  \label{shape-1}
\end{equation}
\begin{equation}
A_{r}=\left(
\begin{array}{ccc}
h_{11}^{r} & h_{12}^{r} & 0 \\
h_{12}^{r} & -h_{11}^{r} & 0 \\
0 & 0 & 0_{n-1}
\end{array}
\right) ,\qquad r=n+3,\ldots ,2m+1.  \label{shape-2}
\end{equation}
\end{th}

{\bf Proof.} In view of the Gauss equation and (\ref{sect-curv}), the scalar curvature
and the mean curvature of $M$ are related by
\begin{equation}
2\tau ={\frac{c}{4}}\left( 3\left\| P\right\| ^{2}+n(n-1)\right) +\left(
n+1\right) ^{2}\left\| H\right\| ^{2}-\left\| h\right\| ^{2},  \label{bi-1}
\end{equation}
where $\left\| P\right\| ^{2}\,$is given by
\[
\left\| P\right\| ^{2}\,=\sum_{i,j=1}^{n+1}g\left( e_{i},Pe_{j}\right) ^{2}
\]
for any local orthonormal basis $\{e_{1},e_{2},\ldots ,e_{n+1}\}$ for $%
T_{p}M $. We introduce
\begin{equation}
\rho =2\tau -\frac{\left( n+1\right) ^{2}\left( n-1\right) }{n}\left\|
H\right\| ^{2}-{\frac{c}{4}}\left( 3\left\| P\right\| ^{2}+n(n-1)\right) .
\label{bi-2}
\end{equation}
From (\ref{bi-1}) and (\ref{bi-2}), we get
\begin{equation}
\left( n+1\right) ^{2}\left\| H\right\| ^{2}=n(\left\| h\right\| ^{2}+\rho ).
\label{bi-3}
\end{equation}
Let $p$ be a point of $M$ and let $\pi \subset T_{p}M$ be a plane section at
$p$. We choose an orthonormal basis $\{e_{1},e_{2},\ldots ,e_{n+1}\}$ for $%
T_{p}M$ and $\{e_{n+2},\ldots ,e_{2m+1}\}$ for the normal space $T_{p}^{\bot
}M$ at $p$ such that $\pi =Span\left\{ e_{1},e_{2}\right\} $ and
the mean curvature vector $H\left( p\right) $ is parallel to $e_{n+2}$, then
from the equation (\ref{bi-3}) we get
\begin{equation}
\left( \sum_{i=1}^{n+1}h_{ii}^{n+2}\right) ^{2}=n\left(
\sum_{i=1}^{n+1}\left( h_{ii}^{n+2}\right) ^{2}+\sum_{i\not{=}j}\left(
h_{ij}^{n+2}\right) ^{2}+\sum_{r={n+3}}^{2m+1}\sum_{i,j=1}^{n+1}\left(
h_{ij}^{r}\right) ^{2}+\rho \right) .  \label{bi-4}
\end{equation}
Using Lemma~\ref{chen-lemma}, from (\ref{bi-4}) we obtain
\begin{equation}
h_{11}^{n+2}h_{22}^{n+2}\geq \frac{1}{2}\left\{ \sum_{i\neq j}\left(
h_{ij}^{n+2}\right) ^{2}+\sum_{r=n+3}^{2m+1}\sum_{i,j=1}^{n+1}\left(
h_{ij}^{r}\right) ^{2}+\rho \right\} .  \label{bi-5}
\end{equation}
From the Gauss equation and (\ref{sect-curv}), we also have
\begin{eqnarray}
K\left( \pi \right) &=&\frac{c}{4}\left( 1+3\alpha \left( \pi \right) -\beta
\left( \pi \right) \right)  \nonumber \\
&&+h_{11}^{n+2}h_{22}^{n+2}-\left( h_{12}^{n+2}\right)
^{2}+\sum_{r=n+3}^{2m+1}\left( h_{11}^{r}h_{22}^{r}-\left( h_{12}^{r}\right)
^{2}\right) .  \label{bi-6}
\end{eqnarray}
Thus, we have
\begin{eqnarray}
K(\pi ) &\geq &\frac{c}{4}\left( 1+3\alpha \left( \pi \right) -\beta \left(
\pi \right) \right) +\frac{1}{2}\rho  \nonumber \\
&&\quad +\sum_{r=n+2}^{2m+1}\sum_{j>2}\{(h_{1j}^{r})^{2}+(h_{2j}^{r})^{2}\}+%
\frac{1}{2}\sum_{i\not{=}j>2}(h_{ij}^{n+2})^{2}  \nonumber \\
&&\quad +\frac{1}{2}\sum_{r=n+3}^{2m+1}\sum_{i,j>2}(h_{ij}^{r})^{2}+\frac{1}{%
2}\sum_{r=n+3}^{2m+1}(h_{11}^{r}+h_{22}^{r})^{2},  \label{bi-7}
\end{eqnarray}
or
\begin{equation}
K\left( \pi \right) \geq \frac{c}{4}\left( 1+3\alpha \left( \pi \right)
-\beta \left( \pi \right) \right) +\frac{1}{2}\rho ,  \label{bi-8}
\end{equation}
which in view of (\ref{bi-2}) yields (\ref{bi}).

If the equality in (\ref{bi}) holds, then the inequalities given by (\ref
{bi-5}) and (\ref{bi-7}) become equalities. In this case, we have
\begin{eqnarray}
h_{1j}^{n+2} &=&0,\ h_{2j}^{n+2}=0,\ h_{ij}^{n+2}=0,\quad i\neq j>2;
\nonumber \\
h_{1j}^{r} &=&h_{2j}^{r}=h_{ij}^{r}=0,\ r=n+3,\ldots ,2m+1;\quad
i,j=3,\ldots ,n+1;  \nonumber \\
h_{11}^{n+3}+h_{22}^{n+3} &=&\cdots =h_{11}^{2m+1}+h_{22}^{2m+1}=0.
\label{bi-9}
\end{eqnarray}
Furthermore, we may choose $e_{1}$ and $e_{2}$ so that $h_{12}^{n+2}=0$.
Moreover, by applying Lemma~\ref{chen-lemma}, we also have
\begin{equation}
h_{11}^{n+2}+h_{22}^{n+2}=h_{33}^{n+2}=\cdots =h_{n+1\ n+1}^{n+2}.
\label{bi-10}
\end{equation}
Thus, choosing a suitable orthonormal basis $\left\{ e_{1},\ldots
,e_{2m+1}\right\} $, the shape operator of $M$ becomes of the form given by (%
\ref{shape-1}) and (\ref{shape-2}). The converse is straightforward.
\medskip

\section{Some applications}

For the case $c=0$, from (\ref{bi}) we have the following pinching result.

\begin{prop}
Let $M$ be an $\left( n+1\right) $-dimensional $\left( n>1\right) $
submanifold isometrically immersed in a $\left( 2m+1\right) $-dimensional
cosymplectic space form $\tilde{M}\left( c\right) $ with $c=0$ such that $%
\xi \in TM$. Then, we have the following
\[
\delta _{M}\leq \frac{\left( n+1\right) ^{2}\left( n-1\right) }{2n}\left\|
H\right\| ^{2}.
\]
\end{prop}

A submanifold $M$ of an almost contact metric manifold $\tilde{M}$ with $\xi
\!\in \!TM$ is called a {\em semi-invariant submanifold}([1])
of $\tilde{M}$ if $TM={\cal D}\oplus {\cal D}^{\perp }\oplus \{\xi \}$,
where ${\cal D}=TM\cap \varphi (TM)$ and ${\cal D}^{\perp }=TM\cap \varphi
(T^{\perp }M)$. In fact, the condition $TM={\cal D}\oplus {\cal D}^{\perp
}\oplus \{\xi \}$ implies that the endomorphism $P$ is an $f${\em -structure}
([12]) on $M$ with $rank\left( P\right) =\dim \left( {\cal D}%
\right) $. A semi-invariant submanifold of an almost contact metric manifold
becomes an {\em invariant }or {\em anti-invariant submanifold} according as
the anti-invariant distribution ${\cal D}^{\perp }$ is $\left\{ 0\right\} $
or invariant distribution ${\cal D}$ is $\left\{ 0\right\} $([1, 12]). \medskip

Now, we establish two inequalities in the following two theorem, which are
analogous to that of (\ref{chen-in-1}).

\begin{th}
\label{th-appl-1}Let $M$ be an $\left( n+1\right) $-dimensional $\left(
n>1\right) $ submanifold isometrically immersed in a $\left( 2m+1\right) $%
-dimensional cosymplectic space form $\tilde{M}\left( c\right) $ such that
the structure vector field $\xi $ is tangential to $M$. If $c<0$, then
\begin{equation}
\delta _{M}\leq \frac{\left( n+1\right) ^{2}\left( n-1\right) }{2n}\left\|
H\right\| ^{2}+\frac{1}{2}\left( n+1\right) \left( n-2\right) \frac{c}{4}.
\label{appl-1}
\end{equation}
The equality in $\left( \ref{appl-1}\right) $ holds if and only if $M$ is a
semi-invariant submanifold with $rank\left( P\right) =2$ and $%
\beta \left( \pi \right) =0$.
\end{th}

{\bf Proof.} Since $c<0$, in order to estimate $\delta _{M}$, we minimize $3\left\|
P\right\| ^{2}-6\alpha (\pi )+2\beta (\pi )$ in (\ref{bi}). For an
orthonormal basis $\left\{ e_{1},\ldots ,e_{n+1}\right\} $ of $T_{p}M$ with $%
\pi =span \left\{ e_{1},e_{2}\right\} $, we write
\[
\left\| P\right\| ^{2}-2\alpha \left( \pi \right) =\sum_{i,j=3}^{n+1}g\left(
e_{i},\varphi e_{j}\right) ^{2}+2\sum_{j=3}^{n+1}\left\{ (g\left(
e_{1},\varphi e_{j}\right) ^{2}+g\left( e_{2},\varphi e_{j}\right)
^{2}\right\} .
\]
Thus, we see that the minimum value of $3\left\| P\right\| ^{2}-6\alpha (\pi
)+2\beta (\pi )$ is zero, provided $\pi =span\left\{
e_{1},e_{2}\right\} $ is orthogonal to $\xi $ and $span\left\{
\varphi e_{j}\ |\ j=3,\cdots ,n\right\} $ is orthogonal to the tangent space
$T_{p}M$. Thus, we have (\ref{appl-1}) with equality case holding if and
only if $M$ is semi-invariant such that $rank\left( P\right) =2$
with $\beta =0$.
\medskip

\begin{th}
\label{th-appl-2}Let $M$ be an $\left( n+1\right) $-dimensional $\left(
n>1\right) $ submanifold isometrically immersed in a $\left( 2m+1\right) $%
-dimensional cosymplectic space form $\tilde{M}\left( c\right) $ such that $%
\xi \in TM$. If $c>0$, then
\begin{equation}
\delta _{M}\leq \frac{\left( n+1\right) ^{2}\left( n-1\right) }{2n}\left\|
H\right\| ^{2}+\frac{1}{2}n\left( n+2\right) \frac{c}{4}.  \label{appl-2}
\end{equation}
The equality in $\left( \ref{appl-2}\right) $ holds if and only if $M$ is an
invariant submanifold and $\beta =1$.
\end{th}

{\bf Proof.} Since $c>0$, in order to estimate $\delta _{M}$, we maximize $3\left\|
P\right\| ^{2}-6\alpha (\pi )+2\beta (\pi )$ in (\ref{bi}). We observe that
the maximum of $3\left\| P\right\| ^{2}-6\alpha (\pi )+2\beta (\pi )$ is
attained for $\left\| P\right\| ^{2}=n$, $\alpha (\pi )=0$ and $\beta (\pi
)=1$, that is, $M$ is invariant and $\xi \in \pi $. Thus, we obtain (\ref
{appl-2}) with equality case if and only if $M$ is invariant with $\beta =1$%
.
\medskip

In last, we prove the following

\begin{th}
\label{th-appl-3}If $M$ is an $\left( n+1\right) $-dimensional $\left(
n>1\right) $ submanifold isometrically immersed in a $\left( 2m+1\right) $%
-dimensional cosymplectic space form $\tilde{M}\left( c\right) $ such that $%
c>0$, $\xi \in TM$ and
\[
\delta _{M}=\frac{\left( n+1\right) ^{2}\left( n-1\right) }{2n}\left\|
H\right\| ^{2}+\frac{1}{2}n\left( n+2\right) \frac{c}{4},
\]
then $M$ is a totally geodesic cosymplectic space form $M\left( c\right) $.
\end{th}

{\bf Proof.} In view of Theorem~\ref{th-appl-2}, $M$ is an odd-dimensional invariant
submanifold of the cosymplectic space form $\tilde{M}(c)$. For every point $%
p\in M$, we can choose an orthonormal basis $\left\{ e_{1}=\xi ,e_{2},\cdots
,e_{n+1}\right\} $ for $T_{p}M$ and $\left\{ e_{n+2},\cdots
,e_{2m+1}\right\} $ for $T_{p}^{\bot }M$ such that $A_{r}$ $(r=n+2,\ldots
,2m+1)$ take the form (\ref{shape-1}) and (\ref{shape-2}). Since $M$ is an
invariant submanifold of a cosymplectic manifold, therefore it is minimal
and $A_{r}\varphi +\varphi A_{r}=0$, $r=n+2,\ldots ,2m+1$([11]).
 Thus all the shape operators take the form
\begin{equation}
A_{r}=\left(
\begin{array}{ccc}
c_{r} & d_{r} & 0 \\
d_{r} & -c_{r} & 0 \\
0 & 0 & 0_{n-1}
\end{array}
\right) ,\qquad r=n+2,\ldots ,2m+1.  \nonumber
\end{equation}
Since, $A_{r}\varphi e_{1}=0$, $r=n+2,\cdots ,2m+1$, from $A_{r}\varphi
+\varphi A_{r}=0$, we get $\varphi A_{r}e_{1}=0$. Applying $\varphi $ to
this equation, we obtain $A_{r}e_{1}=\eta (A_{r}e_{1})\xi =\eta
(A_{r}e_{1})e_{1}$; and thus $d_{r}=0$, $r=n+2,\ldots ,2m+1$. This implies
that $A_{r}e_{2}=-c_{r}e_{2}$. Applying $\varphi $ to the both sides, in
view of $A_{r}\varphi +\varphi A_{r}=0$, we get $A_{r}\varphi
e_{2}=c_{r}\varphi e_{2}$. Since $\varphi e_{2}$ is orthogonal to $\xi $ and
$e_{2}$ and $\varphi $ has maximal rank, the principal curvature $c_{r}$ is
zero. Hence, $M$ becomes totally geodesic. As in Proposition 1.3 on page 313
of [12], it is easy to show that $M$ is a cosymplectic manifold of
constant $\varphi $-sectional curvature $c$.
\medskip

\vspace{0.5cm}
\noindent{\bf Acknowledgment :}
This work was supported by Korea Science \& Engineering Foundation Grant
(R01-2001-00003).

Jeong-Sik Kim

Department of Mathematics Education

Sunchon National University, Sunchon 540-742, Korea

email: jskim01@hanmir.com

\vspace{0.5cm}
Jaedong Choi

Department of Mathematics

P. O. Box 335-2 Airforce Academy

Ssangsu, Namil, Chungwon, Chungbuk,
363-849, Korea

e-mail : jdong@afa.ac.kr

\end{document}